\documentclass[sigconf]{acmart}

\AtBeginDocument{%
  \providecommand\BibTeX{{%
    \normalfont B\kern-0.5em{\scshape i\kern-0.25em b}\kern-0.8em\TeX}}}

\copyrightyear{2021}
\acmYear{2021}
\setcopyright{acmcopyright}\acmConference[ICGJ 2021]{Sixth Annual International Conference on Game Jams, Hackathons, and Game Creation Events}{August 2, 2021}{Montreal, Canada}
\acmBooktitle{Sixth Annual International Conference on Game Jams, Hackathons, and Game Creation Events (ICGJ 2021), August 2, 2021, Montreal, Canada}
\acmPrice{15.00}
\acmDOI{10.1145/3472688.3472689}
\acmISBN{978-1-4503-8417-9/21/08}

\usepackage[utf8]{inputenc}
\usepackage[section]{placeins}
\hypersetup{draft}
\begin{document}

\title{Two Decades of Game Jams}

\author{Gorm Lai}
\email{glai001@gold.ac.uk}
\affiliation{%
  \institution{Goldsmiths, University of London}
  \city{London}
  \country{United Kingdom}
}

\author{Annakaisa Kultima}
\email{annakaisa.kultima@aalto.fi}
\affiliation{%
  \institution{Aalto University}
  \city{Espoo}
  \country{Finland}
}

\author{Foaad Khosmood}
\email{foaad@calpoly.edu}
\affiliation{%
  \institution{Cal Poly}
  \streetaddress{1 Grand Av.}
  \city{San Luis Obispo}
  \state{California}
  \country{USA}}

\author{Johanna Pirker}
\email{johanna.pirker@tugraz.at}
\affiliation{%
  \institution{Graz University of Technology}
  \city{Graz}
  \country{Austria}
}

\author{Allan Fowler}
\email{allan.fowler@auckland.ac.nz}
\affiliation{%
  \institution{University of Auckland}
  \city{Auckland}
  \country{New Zealand}
}

\author{Ilaria Vecchi}
\email{I.Vecchi@leedstrinity.ac.uk}
\affiliation{%
  \institution{Leeds Trinity University}
  \city{Leeds}
  \country{United Kingdom}
}

\author{William Latham}
\email{w.latham@gold.ac.uk}
\affiliation{%
  \institution{Goldsmiths, University of London}
  \city{London}
  \country{United Kingdom}
}

\author{Frederic Fol Leymarie}
\email{f.leymarie@gold.ac.uk}
\affiliation{%
  \institution{Goldsmiths, University of London}
  \city{London}
  \country{United Kingdom}
}

\date{July 2021}

\renewcommand{\shortauthors}{Lai et al.}

\begin{abstract}
In less than a year's time, March 2022 will mark the twentieth anniversary of the first documented game jam, the Indie Game Jam, which took place in Oakland, California in 2002. Initially, game jams were widely seen as frivolous activities. Since then, they have taken the world by storm. Game jams have not only become part of the day-to-day process of many game developers, but jams are also used for activist purposes, for learning and teaching, as part of the experience economy, for making commercial prototypes that gamers can vote on, and more. Beyond only surveying game jams and the relevant published scientific literature from the last two decades, this paper has several additional contributions. It builds a history of game jams, and proposes two different taxonomies of game jams --- a historical and a categorical. In addition, it discusses the definition of game jam and identifies the most active research areas within the game jam community such as the interplay and development with local communities, the study and analysis of game jammers and organisers, and works that bring a critical look on game jams. 
\end{abstract}

%
\maketitle

\section{Introduction}
The game jam movement was born with the inception of the Indie Game Jam in 2002~\cite{Hecker2002}. By March 2022, game jams will officially have existed for 20 years. It is therefore useful to consider the body of work, take a look back at the game jam movement, and discuss how game jams and the study of these have evolved during the last two decades. As we will uncover, the movement and surrounding industry have evolved significantly during those years. From being small events for hobbyists and enthusiasts, game jams are now part of the curriculum of most game-related education programs, as well as part and parcel of the culture of many game companies.\\
In order to be able to study in some depth game jams, we should agree on a definition for what they are. Section~\ref{sec_GameJamDefinition} takes a look at different proposed definitions and discusses them. Then, in section~\ref{sec_AHistoryOfGameJams} we consider the history of the movement, builds a history-based identification of types of game jams, and compare game jams to other rapid-prototyping movements. Following the history-based taxonomy, in section~\ref{sec_TypesOfGameJams} we further develop and expand the game jam taxonomy initially introduced by Fowler et al.~\cite{Fowler2015}. Having so far focused on game jams and events, section~\ref{sec_IdentifiedResearchAreas} looks at game jam focused research and highlights important topics, including: development practices (section~\ref{sec_DevelopmentPractices}), criticisms~\ref{sec_GameJamCriticisms}, group forming (section~\ref{sec_GroupForming}), industry development (section~\ref{sec_IndustryDevelopment}), jammers and their motivations (section~\ref{sec_JammersAndTheirMotivations}), organisers and their motivations (section~\ref{sec_Organisers}), sharing of assets, code, etc. (section~\ref{sec_Sharing}), and  themes used to organise jams (section~\ref{sec_Theme}). Finally we conclude and look towards the next decade of game jam development.

\section{Definition of a Game Jam}
\label{sec_GameJamDefinition}
As evidence of game jams becoming a serious research subject, researchers have long tried to distill a game jam definition. In 2010 Musil et al.~\cite{Musil2010} was first to dissect the game jam concept. Instead of coming up with a single definition, the authors identify seven typical attributes that characterize a game jam, as well as providing a macro and a micro definition of the event type. Essential attributes they identify are time-boxing, a general theme, ad hoc group forming and a communal presentation of all the games. The last attribute shows that the authors do not seem to have been considering online game jams. Additionally, they also describe game jams as typically being hardware and software agnostic. That is something that has become less true with time as sponsored game jams and game jams focusing on specific technologies have become commonplace (section~~\ref{sec_CommercialGameJams} and section~\ref{sec_ChallengeTechnology}). Interestingly, the first game jam, The Indie Game Jam~\cite{Hecker2002}, was not software and hardware agnostic, as all the jammers used a game engine, custom made for the event. Kultima~\cite{Kultima2015} analyzed 20 different papers on game jams and distilled their usage of the term into a single definition ``accelerated, constrained and opportunistic game creation events with public exposure''. The same year, Locke et al.~\cite{Locke2015} took a different perspective with a framework that ``establishes a theoretical basis with which to analyze game jams as disruptive, performative processes that result in original creative artefacts''. A year later, Grace~\cite{Grace2016} argued that it is a mistake to focus on the artifact when differentiating game jams from other time-limited creation events such as hackathons. Instead, the author concludes that a game jam ``emphasizes a state and process'', while hackathons focus on ``a measurable result standardized by a shared sense of competition.''

\subsection{Related Movements}
\label{sec_RelatedMovements}
Game jams, like hackathons, are participatory design events. This is due to the openness of the event, the overarching use of rapid prototyping, and sharing of feedback among the teams during the event and at a show-and-tell that typically occurs at the conclusion of the event. Many game jams also feature follow-up events where non-participants can come and play the games, and talk and provide feedback directly to the creators. 

Game Jams are sometimes described as hackathons~\cite{Borg2020}. As we argue, there is no consensus about such a description. However, it is used by many game jam commmunities. For example, the Global Game Jam (GGJ) website states that ``a \textit{game jam} is essentially a hackathon focused on game development.'' In fact, the appellation game jam or hackathon is often being used interchangeably. For example, Geek Girl Academy~\cite{Kane2014} runs both hackathons (\#SheHacks) and game jams (\#SheHacksGames). The use of similar hashtags indicates a similitude. However, as mentioned before, Grace~\cite{Grace2016} argues that jams focus on state and process, while hackathons focus on results achieved through competition. This schism is reflected in Izvalov et al.'s~\cite{Izvalov2017} when comparing GGJ, the world's largest game jam, and the NASA Space Apps Challenge (NSAC), the world's largest hackathon~\cite{Nasa2012}. The organizers of the hackathon declare the challenge one month before the actual event, emphasizing pre-event activities, while imposing that the local events include an evaluation process. In fact NSAC has three levels of judging that winners can progressively advance through until six global winning teams are chosen in the end. On the other hand, GGJ does not require judging and does not encourage preparatory events. Part of the mystery of game jams is that the theme is not unveiled until the moment the jam starts, emphasizing Grace's~\cite{Grace2016} point that game jams focus on the playful state of the participants. This also correlates with a study by Steinke et al.~\cite{Steinke2016} who found that competitive participants enjoyed the game jamming experience less. Other results from that survey are discussed later in section~\ref{sec_JammersAndTheirMotivations}.

Eberhardt~\cite{Eberhardt2016} elaborates on the many forms and purposes of game jams and how these attract specific types of jammers. In particular, Eberhardt considers the professional game jammer, a specialization of the aforementioned competitive game jammer. There are people who specialize in winning game jams, and their focus will naturally be different than those whose primary purpose of participation is learning, network or fun (the motivations of game jammers is the focus of section~\ref{sec_JammersAndTheirMotivations}). This competitive behaviour jells well with commercial or sponsored game jams, such as the Escape Room Game Jam mentioned by the author. In this instance, according to Grace~\cite{Grace2016}, the game jam starts exhibiting properties more like that of a hackathon, as the focus of the jam changes from the process to building an artifact that can win the jam.

There are also events, which sit in between hackathons and game jams. For example, the Global Gov Jam~\cite{Alencar2018}, is an event focusing on public services, where participants (jammers), public officials and designers produce work as prototypes and learn about a particular policy challenge.

Some game jams started as competitions or ``compos''. For instance Ludum Dare~\cite{Howland2002} was formed in 2002, but its jam mode only appeared in 2010~\cite{Kasprzak2010}. The culture of game development competitions is also closely related to the phenomenon of demoscene, focused on audiovisual computer art~\cite{Reunanen2017}.

\section{A History of Game Jams}
\label{sec_AHistoryOfGameJams}
The history of game jams has been told before, each retelling adding a bit more detail. Recently, Juul~\cite{Juul2019} gave an introduction to the birth of the movement describing the Indie Game Jam~\cite{Hecker2002} and the Nordic Game Jam~\cite{Lai2006}. Chen~\cite{Chen2017} provided a longer overview of the history, while, earlier, Fowler et al.~\cite{Fowler2013} also gave a brief history of the Global Game Jam~\cite{Gold2009}. Here we augment that historical knowledge with information about several other game jams, such as the Lithuanian Game Developers Jam-Session~\cite{Pranckevicius2002} (LT Game Jam), one of the earliest documented game jams in the world, only trailing the 0th Indie Game Jam by a few months. We use the historic information to build a series of game jam generation or \emph{waves} of game jams. The appearance of a new wave, does not imply that previous waves die out. The first recorded event in the world, where the term \emph{game jam} was used, was the 0th Indie Game Jam~\cite{Hecker2002}, which took place on 15th-18th of March 2002.

\subsubsection*{Prejams} Rapid prototyping and time-limited challenge events existed before game jams. For example, NaNoWriMo~\cite{NaNoWriMo1999} started as a writing challenge already in 1999. The Interactive Fiction Competition~\cite{Nelson1995}, which was initiated in 1995, describes nowadays submissions as games on the event website, even though the term and concept of \emph{game jam} was established later.

\subsubsection*{Game jams as niche/personal spaces} Ludum Dare~\cite{Howland2002}, the first recorded online-only game jam, began in 2002. The first generation of game jams were tech and programmer focused, and most games were developed by a single person. This is seen in how both the Indie Game Jam~\cite{Hecker2002} and LT Game Jam~\cite{Pranckevicius2002} mandated the use of a specified custom made engine, which would require a programmer. Half the games at Indie Game Jam 0 and three out of ten games LT Game Jam were made by a single person. The LT Game Jam website states that everyone at the game jam were programmers. Originally, Ludum Dare did not call itself a game jam but an ``accelerated solo game development competition'', and so mandated that all games were made by a single person. Currently, Ludum Dare covers two different types of events, a jam and a competition. The latter still requires that the game is made by a single person.
\begin{table}[!htb]
\centering
\begin{tabular}{|p{1.0cm}|p{1.0cm}|p{3.9cm}|p{0.7cm}|} 
  \hline
 \textit{1st Year} & \textit{Month} & \textit{Game Jam} & \textit{Wave} \\
 \hline
 \hline
 2002 & March & Indie Game Jam~\cite{Hecker2002} & 1st \\
 2002 & April & Ludum Dare~\cite{Howland2002} & 1st \\
 2002 & August & LT Game Jam~\cite{Pranckevicius2002} & 1st \\
 \hline
 2006 & January & Nordic Game Jam~\cite{Lai2006} & 2nd\\
 2006 & May & Toronto Game Jam~\cite{McGinley2006} & 2nd\\
 2008 & July & No More Sweden~\cite{Svedang2008} & 2nd\\ 
 \hline
 2009 & January & Global Game Jam~\cite{Gold2009} & 3rd\\
 2010 & May & Health Games Challenge~\cite{Shin2012} & 3rd\\
 2011 & August & Fukushima Game Jam~\cite{Shin2012} & 3rd\\
 \hline
 2013 & March & Train Jam~\cite{Gold2009} & 4th\\
 2015 & & Castle Game Jam~\cite{Newnorth2015} & 4th\\
 \hline
\end{tabular}
\caption{A selection of Impactful Game Jams}
\label{table_gamejamoverview}
\vspace{-8mm}
\end{table}

\subsubsection*{Game Jams as regional communal spaces} Game jams that were part of the second wave actively encouraged team creation. The ToJam~\cite{McGinley2006} invited floating graphics artists and sound designers to support teams that lacked those skills, while the Nordic Game Jam~\cite{Lai2006} organised ice breakers for the jammers, and pitching sessions with group forming (section~\ref{sec_GroupForming}). The second wave of game jams began four years after the 0th Indie Game Jam~\cite{Hecker2002}, as the two game jams that would be competitors for the world's largest single-site game jam launched: The Nordic Game Jam in January 2006~\cite{Lai2006} and the Toronto Game Jam~\cite{McGinley2006} in May 2006. The model that the Nordic Game Jam has set up for organizing a game jam, with its agenda of ice breaker, theme reveal, pitching session and group forming before entering into approximately 48 hours of jamming followed by a final show-and-tell of the games, is the model often referred to by game jam guides such as the Game Jam Guide~\cite{Cornish2012} and popularized by GGJ. This original model of a theme reveal, followed by a pitching session and group forming, is what H{\o}jsted et al.~\cite{Hojsted2010} refer to as the ``Capitalist Method''. This term and game jam model is discussed further in section~\ref{sec_GroupForming}. As a co-founder of the Nordic Game Jam, Juul~\cite{Juul2019} writes about this transition from single developer to a team-centric collaborative event. No More Sweden\footnote{The first No More Sweden, like Ludum Dare, initially described itself as a game competition, however, already in the second year, the term jam was used~\cite{Svedang2009}.} is another early impactful 2nd wave game jam, that still exists.

\subsubsection*{Internationally synchronized game jams} The third wave heralded the advent of distributed game jams with physical locations as well as purposeful game jams~(section~\ref{sec_PurposefulGameJams}). Fukushima Game Jam~\cite{Shin2012} is an example of a purposeful distributed game jam with multiple physical locations. Just as GGJ was modelled upon the Nordic Game Jam, the former inspired additional distributed game jams with many physical locations. In 2012, Shin et al.~\cite{Shin2012} discuss the concept of the \emph{localised} GGJ, which applies the core GGJ model on a regional or nationwide scale. Examples are the Health Games Challenge in the US in 2010, and the Fukushima Game Jam~\cite{Shin2012} in 2012, which was started after the horrific aftermath of a big earthquake and tsunami off the coast of Tohoku, Japan.

\subsubsection*{Game jams as part of the experience economy} A fourth wave introduced game jams into the experience economy~\cite{Pine2011} starting with the Train Jam~\cite{Wallick2013}. This is discussed in much more detail in section~\ref{sec_JamsWhileMoving}. Table~\ref{table_gamejamoverview} shows a selection of impactful game jams belonging to each generation.

While the 3rd and 4th generation of jams do introduce new distinctive elements, already at this stage it becomes difficult to keep track of all the different types of game jams which start appearing. Today, game jams have become popular to a degree that would be unfathomable in earlier years and any sort of categorization becomes a much more complex task. For example, by scraping itch.io~\cite{Corcoran2013}, a website that launched in 2013, Vu \& Bezemer~\cite{Vu2020} analysed 1290 game jams organised on that platform alone. IndieGameJams.com~\cite{Hadley2020} is a website for tracking game jams. 
\section{Taxonomy}
\label{sec_TypesOfGameJams}
Fowler et al.~\cite{Fowler2015} provide a taxonomy of game jams, dividing them into six groups; regional affiliation, setting, technology, career advancement, social/charitable topics and challenge. The taxonomy given here builds and expands upon that. In some cases, we have kept or renamed the original category, while in other cases, we have subdivided them into more groups. We have identified more categories than they did, and changed the content of some that they previously identified. The \emph{regional affiliation} group is the only one we have kept as is, although we have expanded it with more examples. For the \emph{settings} category we have put most of the jams mentioned by Fowler et al~\cite{Fowler2015} into the experience economy group~(section~\ref{sec_JamsWhileMoving}), but some, such as the Phoenix Comicon Game Jam, we instead put into what we call ``part of other events''~(section~\ref{sec_PartOfOtherEvents}). Table~\ref{table_taxonomymapping} gives an overview of the mapping between our taxonomy and the one presented by Fowler et al.~\cite{Fowler2015}. It is important to note that this is a one-way mapping. For example, while all the games jams presented by Fowler et al.~\cite{Fowler2015} under Career Advancement fall into our category of Commercial Game Jams, the reverse is not true. Amnesia Fortnight is a commercial game jam, but classifying it solely as an internal game jam that is meant for career advancement would not reveal its dual purpose of justice~(section~\ref{sec_CommercialGameJams}).

\begin{table}[!htb]
\centering
\begin{tabular}{|p{3.5cm}|p{3.5cm}|} 
  \hline
 \textit{Fowler et al.~\cite{Fowler2015}} & \textit{Our taxonomy} \\
 \hline
 \hline
 & Games Industry Commentary, Meta \\
 \hline
Career Advancement & Commercial Game Jams \\
 \hline
Challenge & Challenge \\
 \hline
Regional Affiliation & Regional Affiliation \\
 \hline
Setting & Experience Economy, Part of Other Events \\
 \hline
Social / Charitable topics & Purposeful Game Jams, Teaching \& Learning \\
 \hline
Technology & Technology \\
 \hline
\end{tabular}
\caption{Mapping taxonomies}
\label{table_taxonomymapping}
\vspace{-8mm}
\end{table}
\subsection{Challenge \& Technology}
\label{sec_ChallengeTechnology}

Referring to Fowler et al.~\cite{Fowler2015} for examples of game jams that fit into these categories, we can also ask the question if events in the categories are even suitable to be categorized as game jams. The focus on technology and the exploration of a specific API or SDK is reminiscent of hackathons --- an activity related to game jams (section~\ref{sec_RelatedMovements}). Looking at the challenge-based game jams, Grace~\cite{Grace2016} argues that game jams focus on the process and not the artifact, leading us to the conclusion that highly competitive game creation competitions, might not necessarily be categorised as game jams. Ludum Dare~\cite{Howland2002} supports this perspective with its division of its game events into jams and competitions. As most game jams have some element of friendly competition, exactly how competitive a game jam needs to be, to take enough focus away from the process of creation to no longer be called a jam is an open question. As described in section~\ref{sec_GameJamDefinition}, with Kultima's~\cite{Kultima2015} definition, this category would still fall under game jams.

\subsection{Commercial Game Jams}
\label{sec_CommercialGameJams}
Double Fine's Amnesia Fortnight was the first game jam to showcase the commercial potential of game jams~\cite{DoubleFineProductions2012}. Starting in 2012, the game jam saw employees at Double Fine  pitch their ideas online. People could then pay to vote for their favourite idea, after which teams at Double Fine spent the next 2 weeks, making the top 5 ideas into playable prototypes. These prototypes could then be downloaded for a fee. With the help of 2 Player Productions, Double Fine also produced several documentaries on Amnesia Fortnight~\cite{DoubleFineProductions2016, DoubleFineProductions2017, DoubleFineProductions2021} as the event has repeated itself through the years.

Game jams for which the entire raison d'{\^e}tre is to promote a product or company have appeared in recent years. Examples are the Epic MegaJam~\cite{EpicGames2020}, Google Stadia Play Anywhere Game Jam~\cite{Google2021} and Oculus' VR Jam~\cite{Oculus2013}.

\subsection{Experience Economy}
\label{sec_JamsWhileMoving}
Starting with Train Jam, a special type of game jam has emerged, where jammers physically travel together while jamming. Train Jam~\cite{Wallick2013}, which took place for the first time in 2013, was the first major jam of this kind. At Train Jam, jammers travel on the AmTrak California Zephyr train, for a scheduled 52 hours, from Chicago to Emeryville, California, arriving just in time for the annual Game Developers Conference to begin in nearby San Francisco. Jamming on a train offers its own rewards and challenges, as wireless connectivity is scarce, jammers must co-exist in a relatively small space for the duration of the jam. The trip offers amazing views of the American countryside and small towns where the train stops. LocomoJam~\cite{Walker2019} and Amaze Train Jam~\cite{Kastner2020} offer similar experiences, the former in Australia going from Brisbane via Sydney to Melbourne~\cite{Walker2019}, and the latter in South Africa from Johannesburg to Cape Town~\cite{Kastner2020}. Organizers also came up with variations on the means of travel, as jammers took to the sea (Splash Jam~\cite{Myrlund2016}, Pirate Jam~\cite{McGee2017}) and air (GDC Plane Jam~\cite{Atkinson-Jones2013}). Table~\ref{table_TravellingGameJams} lists different moving game jams. Kultima~\cite{Kultima2019} reports on several moving game jams which have taken place in Finland, such as the Bus Jam (jam on a bus) and JamBike17~\cite{Zhamul2017} (a two person jam taking 600km journey on a tandem bike).

Moving jams offer unique experiences, with jammers confined to a limited area while being unable to leave, limited connectivity and plenty of other challenges while being provided new experiences. Under adverse conditions, jammers manage to create a game, meet new friends and undertake an unforgettable journey. Game Jams have for certain joined the \emph{Experience Economy}~\cite{Pine2011}. While there has been plenty of writing in the media on moving game jams such as the Train Jam~\cite{Wallick2013}, no academic studies have so far been published on this topic. 

In a similar fashion to travelling game jams, game jams at unique and exciting locations have emerged. The first of these was the Swiss Castle Game Jam~\cite{GGJSuisse2014}, but it was quickly joined by others such as the Castle Jam~\cite{Newnorth2015}, Exile~\cite{Garbos2011}\footnote{On the website of the Exile Game Jam (http://exile.dk) it says that Exile is not a game jam. However, it follows much of the usual format of a game jam, and one of the first posters produced referred to it as \emph{Exile Indie Game Jam}.}, Isolation Game Jam, Quantum Game Jam (two of them in planetariums and one on a Ferris wheel) and Survival Mode~\cite{Kultima2016} which took place in a small cabin in the Finnish Lapland. The cabin had no running water, and there was only limited electricity and internet connection provided through cell phones. Kultima et al.~\cite{Kultima2016} interviewed the participants about their experiences at the Survival Mode and provided an interesting discussion about the constraints experienced at that game jam and game jams in general, versus the constraints of commercial game development.

\begin{table}[!htb]
\centering
\begin{tabular}{|p{2.0cm}|p{3.0cm}|p{2.0cm}|} 
  \hline
 \textit{Founding Year} & \textit{Game Jam} & \textit{Means of Travel} \\
 \hline
 \hline
 2013 & Train Jam~\cite{Wallick2013} & Train \\
 2013 & GDC Plane Jam~\cite{Atkinson-Jones2013} & Airplane \\
 2016 & Splash Jam~\cite{Myrlund2016} & Cruise Ship \\
 2017 & JamBike17~\cite{Zhamul2017} & Tandem Bike \\
 2017 & Pirate Jam~\cite{McGee2017} & Sail boat \\
 2019 & AMAZE Train Jam~\cite{Kastner2020} & Train \\
 2019 & LocomoJam~\cite{Walker2019} & Train \\
 \hline
\end{tabular}
\caption{Overview of travelling game jams}
\label{table_TravellingGameJams}
\vspace{-9mm}
\end{table}

\subsection{Games Industry Commentary}
Game jams have also been used for activism and rhetorical commentary by games industry insiders about happenings inside the games industry. Examples of this include Candy Jam~\cite{Raymond2014}, Flappy Jam~\cite{Somepx2014} and Moly Jam~\cite{Kipnis2012}. Candy Jam was a reaction to the developer Kings' complaint about the registration of the trademark Banner Saga, as King already had a game with the word Saga in it, namely Candy Crash Saga. Candy Jam was created as a satiric reaction to this protectionism, and the creators were encouraged to develop games involving candy. Moly Jam was born out of a Twitter account parodying the game designer Peter Molyneux, who, even in his own words, is famous for over-promising. Flappy Bird was a game that made Vietnamese game designer Dong Nguyen famous almost overnight. In the words of Juul~\cite{Juul2019} ``Because the Vietnamese developer Dong Nguyen did not fit the template for an independent game developer, much of the video game press decided to present Flappy Bird as a cynical copy of older games, rather than as an interesting experimental game paying homage to the past''.  When Nguyen pulled his game from the Apple app store because of what he said was getting too much attention, Flappy Jam~\cite{Somepx2014} was created as a celebration of the game.

\subsection{Meta}
Some game jams play with the definition of game jams and the foundational elements, for example the time limitation. Often the time limit is short enough that the activity is time-boxed, and we fall within the definition of a game jam, but not so short that it becomes an extraordinary constraint. However, some game jams like the 0h Game Jam~\cite{Sosowski2011} last precisely 1 hour, from 2 AM to 2 AM when European countries change from summertime to wintertime. Powley et al.~\cite{Powley2017} also mention that their tool is so easy to use that it can be used for lunchtime game jams.

\subsection{Part Of Other Events}
\label{sec_PartOfOtherEvents}
It has become commonplace for events of all kinds, which are normally related to games jams or in some cases, even games, to host game jams as events within an event. Starting with academia, events such as the Schloss Dagstuhl Seminars~\cite{LeibnizZentrum1990} have had a game jam attended by game AI researchers. This resulted in a paper arguing why game jams should be held at academic conferences, emphasizing that game jams provide opportunity for collaboration and ``playable research''~\cite{Cook2015}.
Jams are also run as part of games industry conferences and exhibitions. Examples are Develop~\cite{Develop2021}, EGX Rezzed~\cite{CreativeAssembly2013}, Pocket Gamer Connects Helsinki~\cite{PocketGamer2021}, GamesCom~\cite{InnoGames2016}. Game jams as events inside events, have also expanded beyond games, for example at Ars Electronica~\cite{ArsElectronica2013, Hirsch2017}. ASM Game Jam~\cite{FinnishGameJam2013}, run as part of the computer culture festival Assembly Summer Festival (originally a demoparty).

\subsection{Purposeful Game Jams}
\label{sec_PurposefulGameJams}
With climate-change being one of the themes of our time, there have been several incarnations of climate game jams. put on by governments~\cite{Gov2015}, universities~\cite{Foltz2019} and private organisations~\cite{Indiecade2020}. Related to game jams for serious purposes, in recent years, game jam organisers have taken the idea one step further and used jams as part of an activist movement. Examples of activist games jam include Resist Jam~\cite{Wallin2017}, Anti-Fascist Game Jam~\cite{GameCurious2020}, Lyst~\cite{CopenhagenGameCollective2014} and ADL's Being An Ally Game Jam~\cite{ADL2017}. Specifically, as seen with Boob Jam~\cite{Frank2013}, XX Game Jam~\cite{Kennedy2018}, GCON's game jams~\cite{Salim2012, GBiz2020}, Women Game Jam~\cite{WGJ2018} and Myerscough et al.'s use of game jams to promote inclusivity~\cite{Myerscough2017}, these examples show that game jams have often been used as diversity incubators and as a way to address the lack of diverse representation in the games- and wider tech-industry.

Game jams have been used as means for learning about, interacting with, and reflecting on indigenous cultures such the Sami, as highlighted by Kultima \& Laiti~\cite{Kultima2019a} and Laiti et al.~\cite{Laiti2020}. In a similar approach, Guo et al.~\cite{Guo2020} used game jams as a tool for examining national heritage.

The Fukushima Game Jam~\cite{Shin2012} was started as a disaster recovery effort in the aftermath of the Fukushima nuclear disaster, and has since evolved to become an annual event in Japan.

News Jam~\cite{Grace2018} set out to prove that ``games can be produced at the pace of news cycles''. During the News Jam, which took place in two locations simultaneously, 24 jammers created five games with a focus on events that were recent when the jam took place, such as the California wildfires and the hurricane in Puerto Rice. At least one of the \emph{news games} received news coverage by traditional international news outlets.
\subsection{Regional Affiliation}
\label{sec_RegionalAffiliation}
Fowler et al.~\cite{Fowler2015} list a number of regional game jams: ``Global Game Jam~\cite{Gold2009}, Nordic Game Jam~\cite{Lai2006}, ToJam~\cite{McGinley2006}, Scottish Game Jam~\cite{McDonald2015}, Finnish Game Jam~\cite{Kultima2010}, Kiwi Jam~\cite{Kenobi2013}''. Here we will add Slavic Game Jam~\cite{KNTGPolygon2015} and No More Sweden~\cite{Svedang2008} to that list. Note that some of these, such as the Nordic Game Jam, represent singular events, while others, such as the Finnish Game Jam, represent regional movements that organise game jams.

\subsection{Teaching \& Learning}
Several authors have focused on using game jams for promoting STEM and STEAM education. Fowler~\cite{Fowler2016} argues for using game jams, hackathons and maker spaces as informal learning environments for STEM education. Fowler \& Schreiber~\cite{Fowler2017} created a game jam summer camp framework designed to encourage people from under-represented and minority backgrounds to get a STEM education. Similarly, Pollock et al.~\cite{Pollock2017} combined game making and neuroscience into after school programs to entice young people to get a STEAM education. The Indie Galactic Space Jam~\cite{Indienomicon2014} is an annual game jam about space travel.

Based on the concept of Critical Pedagogy, Myers et al.~\cite{Myers2019} build a framework for game jams designed to democratize knowledge, explore and teach the participants about a social issue. The focus is not the artefact, but on making the jammers into ``agents of social change.'' Applying their own framework, the authors create a game jam focusing on teaching the participants about everyday sexism. 

Meril{\"{a}}inen et al.~\cite{Merilainen2020} recently published a survey of the use of game jams in learning environments and found three emergent themes: game development skills, STE(A)M skills and personal \& interpersonal skills. The first theme focused on formal and informal learning of skills in preparation for a career in game development. The STE(A)M theme focused on using game jams to promote STEM \& STEAM, in general, and to attract under-represented groups to the included fields. The authors noted that many papers mentioned that while harder to measure, game jams are also suitable for teaching social skills. However, the authors also say that this is a potential benefit that needs more examination, as the papers that do mention this benefit offer little evidence. The survey also looks further into the challenges of all three themes and discusses their findings.

In 2013 the UK's Royal Society organised a summer of science game jam~\cite{RoyalSociety2013}, where jammers paired up with scientists to produce games that conveyed scientific ideas.

\section{Research Areas}
Surveying the literature on game jams, we have identified a number of research topics such as development practices, game jam criticisms, group forming and jammers \& their motivations. In the following, we review works falling into each category. 
\label{sec_IdentifiedResearchAreas}
\subsection{Development Practices}
\label{sec_DevelopmentPractices}
Game jams typically use a central theme and other constraints, such as GGJ's diversifiers, in part to create an even playing field and spark the jammers' imagination. However, additional methods such as ideation methods and toolkits can be used to create more ideas, spark discussions and flesh out details. Ho~\cite{Ho2017} surveyed a list of toolkits and websites designed for ideation.

Mechanics Dynamics Aesthetics (MDA) is a game design framework by Hunicke et al.~\cite{Hunicke2004} that places itself in the middle between different development models such as waterfall and an iterative bottom-up approach. It starts with a focus on aesthetics --- what do we want the player to feel? The method encourages iteration as the focus moves between the layers. Buttfield-Addison et al.~\cite{Buttfield-Addison2016} examined the idea of applying MDA to the development process at game jams. Game jams are as much about production as they are about design, and the authors argue that it is rare to see MDA applied in this context. The authors' argument is that after choosing a mutual base of mechanics and aesthetics, it is possible for the team to apply production roles to the different MDA layers at a game jam: programmers focus on mechanics, game designers on dynamics and artists on the aesthetics. As the jam and game production progresses, communication between these layers is needed.

Farhan \& Kocher~\cite{Farhan2016} write about attempts in the Swiss games industry to use game jams for emulating the big teams typical for big game productions. They describe two such attempts, $I^{3}$ and the Swiss Mercenaries Workshop. The jams had different formats. $I^{3}$ existed in four various iterations, with 26 participants. Unlike most other game jams described here  $I^{3}$ was invitational only. It was organised by a single person who also acted as the producer on the project. Where  $I^{3}$ lasted between one and three days, the different iterations of the Swiss Mercenaries Workshop lasted for three weeks, with a lead team doing the main preparations during those weeks and the rests joining for the main event. Where the  $I^{3}$ stood for III (as opposed to AAA) and was a community event, the Swiss Mercenaries Workshop was targeted at students as way to learn how to work in big teams. The Big Team game jams are interesting as they can work as learning experiences and replacement experiences for regions with few big companies. On the other hand, they also represent a break away from typical game jam experiences as they are selective about participants and focus on emulating some of the structures and hierarchy of game companies. Hierarchy and selection based on participants' skill level are not attributes that are typically emulated as part of the group dynamics at game jams.

\subsection{Game Jam Criticisms}
\label{sec_GameJamCriticisms}
As game jams became more popular, criticisms of game jams started to appear from established games developers. These have primarily focused on the idea that a successful game jam prototype can easily be turned into a successful, more fully-fledged game. In his talk at No More Sweden 2010, indie developer Michael Todd spoke about the difficulty of transforming smaller games into bigger ones~\cite{Todd2010}. Chris Hecker, one of the founders of the first documented game jam~\cite{Hecker2002}, had a talk at the Game Developers Conference 2010~\cite{Informa1988} entitled "Please Finish Your Games", and then in 2012 co-organised a retreat, the Depth Jam, where the four participants focused on improving details of existing games. The implication here is that \emph{regular} game jam games end being shallow and horizontal in the way they explore their mechanics. Both Hecker and Blow, who co-organised the Depth Jam, have written extensive blog pieces with criticism of modern game jams and their motivations for organising the Depth Jam~\cite{Hecker2012, Blow2012}. In 2013, at GDC Next, Nathan Vella spoke about the challenges of making game jam games into full games~\cite{Vella2013} and the difference in mindset needed in the two settings. For example, Vella suggests that jamming and prototyping, two skills we often conflate, are different. The former is used at game jams, while the latter is used for iterating on existing projects.

Fullerton et al.~\cite{Fullerton2006} in one of the first papers on the subject, say that game jams are ``a fun way to innovate, to be sure, but only a few of the games produced during these highly energized events have provided inklings of true innovation, and unfortunately, none have seen any application past their initial demonstrations in respective showcases''. That quote is not as true as it used to be, as a number of games which started as game jam prototypes, have since gone onto commercial and critical success~\cite{GGJSuccess2017}. However, success is rare, and proponents of game jams often emphasize other features of game jams such as learning and networking (section~\ref{sec_JammersAndTheirMotivations}).

In their survey on GGJ participants, Borg et al.~\cite{Borg2020} connect crunching, which is prevalent in the games industry~\cite{Edholm2017}, with the intensity of a 48-hour game jam, where some participants replicate game industry behaviour by sacrificing sleep and recuperation for working on their prototype. Participants in the survey by Borg et al.~\cite{Borg2020} describe how they willingly crunch to get as much done during the game jam as they can. In recent years, GGJ has become focused on directly discouraging crunch, for example through the keynote~\cite{GGJ2011} and focus on the mental health of the jammers~\cite{GGJ2018, GGJ2020b}. Several other authors mention jammers crunching~\cite{Reng2013, Faas2019}.

While game jams have found their way into many different areas, it is not straightforward to apply the game jam model in any context, as evidenced by Yamane~\cite{Yamane2013}. They document that participatory design theory have had trouble gaining traction in Japan. However, in 2010 the same year as the first GGJ in Japan, the idea of rapid prototyping took a foothold, as game developers working on Final Fantasy XIII documented the usefulness of the approach.

In a Marxist analysis of game jams, Rossi~\cite{Rossi2019} identifies game jammers as members of the cognitariat, who with their enthusiasm provide free labour and for whom the division between work and free time become blurred.

\subsection{Group Forming}
\label{sec_GroupForming}
The Global Game Jam~\cite{Gold2009} focuses on teams making games and group forming, a feature inherited from the Nordic Game Jam~\cite{Lai2006}. This is a way to ensure that jammers have a good experience participating in GGJ and that they'd like to return. The Nordic Game Jam pioneered idea pitching as a basis for group forming at game jams. In this setup, jammers pitch their ideas at the beginning of the jam, and jammers rally and form teams around the ideas they like the best. H{\o}jsted et al.~\cite{Hojsted2010} call this the Capitalist Method. Opposed to this is the Socialist Method, which focuses on forming the groups before the game concept is ideated. Thus the main conceptual difference between the Capitalist or the Socialist methods is that the former is \emph{idea first}, while the latter is \emph{group first}.  H{\o}jsted et al.~\cite{Hojsted2010} describe the Socialist approach as a more scalable group forming method than the pitching (Capitalist) approach. Studies on group forming have so far mainly focused on pitching-first methods.

Pirker \& Voll~\cite{Pirker2015} tested jammer satisfaction with three methods, all based on the pitching method: unsupervised, supervised intervention based on skills, and supervised intervention based on skills and tech/engine. Their study showed that while all group forming methods led to relatively high satisfaction, supervised methods rated higher than non-supervised ones, and the method with the most intervention rated the highest. These methods were all tested at game jams with a small number of participants. Pirker \& Voll also describe some differences in the team forming methods at a smaller jam (Game Jam Graz~\cite{Pirker2014}) versus the larger Vancouver GGJ~\cite{Voll2015}, which housed over 350 participants in 2015. GGJ Vancouver allows jammers to come with pre-formed teams and facilitate pitching sessions for the remainder. Also, jammers supporting several teams (such as artists and especially sound designers) are an important element to consider. Like many other jams, GGJ Vancouver centralised audio as a service, gathering sound designers in one group that then out-source their skills to the other game-making groups because of the specific nature of the job.

\subsection{Industry Development}
\label{sec_IndustryDevelopment}
In popular jargon, one of the transformations that the games industry has undergone from the mid 2000s has been the \emph{democratization of game development}. Proceeding without discussing the correctness of the semantics of this phrase, the concept has come to mean that game-development, -production and -publishing has become widely accessible. For example, though independent game development has existed since the birth of the games industry, self-publishing on consoles and existing game platforms was almost inaccessible until the mid-2000s~\citep{Juul2019}. Having had their conceptual birth in the 2000s, game jams have played a significant role in this democratization process, as many successful games have had their initial prototype made at a game jam and makers of game engines, such as Unity and Unreal have become major sponsors of game jams. Another point made by Musil et al~\cite{Musil2010} is that the smaller apps typical for the handheld devices fit game jams well, as these events focus on experiences that can quickly be prototyped in a small amount of time. In a comparison to how some punk bands were made in the 1970s and 1980s, Sampugnaro et al.~\cite{Sampugnaro2014} describe game jammers as taking part in punk capitalism, and go on to say ``that the birth and growth of alternative channels of production and distribution proves game development as the leader of this revolution.''

Izvalov et al.~\cite{Izvalov2016} details how GGJ was used as a way to develop the Ukrainian IT industry. They describe how GGJ brought together technical and creative companies and how the reach of GGJ has made it easier to start and attract other IT events. Surveying participants at a game jam and two game development projects, Pirker et al.~\cite{Pirker2016} argues for the benefits of jamming and rapid prototyping not only for developing STEM skills, but also for developing other industry-relevant skills, building a portfolio, and building a social network. They also argue that game jams are a great place for industry people to meet young developers.

\subsection{Jammers \& Their Motivations}
\label{sec_JammersAndTheirMotivations}
As emphasized by Kennedy~\cite{Kennedy2018} and evident from the pictures on the Indie Game Jam website, game jams started out mostly as a male-oriented activity. As we will see, just like the general demography of the games industry has changed, so has that of game jams. Several works have examined the make up of game jams. Fowler et al.~\cite{Fowler2013} describe the evolution of the GGJ through the years 2009-2013. We can study game jam motivations from at least two different perspectives: the game jammer and the game jam organizer. Most studies, like this section, focus on the former, while section~\ref{sec_Organisers} looks at studies focusing on organizers.

Part of Zook \& Riedl's study on the game development process~\cite{Zook2013} of jammers at the 2013 GGJ asked about the goals of the game design while making a game. Examining a mechanic was the most common answer. Other answers included learning a skill, meeting people and raising awareness. This fits well with a study by Batista et al.~\cite{Melo2019}, who found that while gaining technical knowledge is important for students and hobbyists, indies and professional game developers value social and personal motivations for participating. Just like Zook \& Riedl, Sampugnaro et al~\cite{Sampugnaro2014} ran a study of 2013 GGJ. Looking at educational backgrounds, with numbers averaged from across the world, 31\% of jammers reported having a bachelor degree as the highest degree, while 10\% reported having a masters degree. This matches the numbers in the US somewhat for 2018~\cite{Wikipedia_USDegrees}, where circa 35\% reported holding a bachelor degree, and circa 13\% a master degree. It would be interesting to see how these numbers compare on a global scale. Of course, the game jam demography could also have changed substantially in the five years between 2013 and 2018. so a new study would have to take this into account. Their survey also reveals that, looking at the college and universities studies of the jammers, only 10.5\% has a background in game development. Their study also reveals that 12.6\% of the attendees identified as women.

Focusing on the social skills of jammers, Smith \& Bowers~\cite{Smith2016} found that game jams improve the self-efficacy of jammers. By comparing pre-and post-jam surveys returned by the jammers, the authors conclude that jammers are significantly more likely to make friends, get better at reaching out to others and get them to help, as well as influence their team, communicate their ideas, and so forth. 

Examining responses from the GGJ annual jammer survey from 2013-2016,  Steinke et al. ~\cite{Steinke2016} found that jammers are mainly young adults. More than 40\% of the jammers are between the ages of 18 and 24. Except for one case, jammers residing in Asia in 2015, jammers between 31 and 40 make up less than 20\% of the participants. There are older jammers participating, but they make up a small group. They also find that motivations for participating in GGJ correlate with age. Networking and having fun increase in importance as the jammers age, while learning decreases as a reason to participate. Intuitively, in addition, they found that experienced jammers have a better jam experience. On the other hand, it is expected that experienced jammers know how to better customize the jam experience to their personal needs and wants, and on the other hand, it might be that participants who do not enjoy jamming simply stop going.

Through informal talks with experienced jammers at GGJ MIT, Eberhardt~\cite{Eberhardt2016} learned that often their primary motivation is to finish a game. This motivation can lead to a bad experience for teams consisting of both novice and experienced jammers.

Kultima focused on Superjammers~\cite{Kultima2019}, whom she defines as game jammers who have participated in more than 20 game jams, and interviewed six such people from the Finnish game jam scene. These jammers repeated many of the same reasons as other jammers~\cite{Zook2013} such as learning new skills and networking. While these super jammers exist and are often highly profiled, analyzing data from GGJ 2014-2016, Pirker et al.~\cite{Pirker2017} used Social Network Analysis (SNA) to map the social network between jammers, and it turns out that most jammers barely move between locations. The authors argue that instead of creating one large graph, the social network is divided into subgraphs. On average, jammers have 4.335 connections to other jammers, and few have over nine connections. Over 1500 jammers have a connection degree of 1. Their research also shows that jammers are likely to have worked with other jammers they have previously worked with. Pirker et al. expanded this work with an even deeper social network analysis~\cite{Pirker2018}. The work also demonstrates that jammers with skills such as audio and art tend to have a higher degree compared to programmers for instance. This shows that these jammers tend to change groups more often and are often involved in more than one group per jam. 

While much work has focused on the jammers, non-jammers have gotten little focus. Wearn et al.~\cite{Wearn2014} set out to find out why some of the students at their university do not attend GGJ. The authors work at a university where nearly 700 students had some kind of game development as part of their degree, though only 297 of them signed up for GGJ 2014. This is especially interesting, given some of the documented benefits of game jams, such as networking, finishing a game, learning new skills and having fun. Using a survey, many of the responses indicate a lack of awareness of the event, though some also purposefully avoided the event with reasons such as time attending these jams ``would be better spent working on actual projects'', ``my previous experience of the jam was that it disrupted my sleep and buzz for the coming semester'' and ``have done several game jams now and am more interested in a longer-term project outside the scope of a game jam''. The authors start the article by stating that they want to look into reasons connected with race, gender or academic achievement. Unfortunately, if they had any results on these factors as they relate to GGJ attendance, they do not elaborate upon them in the paper. Meril{\"{a}}inen \& Aurava~\cite{Merilainen2018} studied reasons for non-participation in game jams. Through interviews with participants at GGJ~\cite{Gold2009}, they uncovered four themes that could be demotivating for participation, including pre-conceptions about game jams, participants and personal insecurities. They also identify initiatives that organisers can take in order to ease the anxiety of first-time participants.

Wearn \& McDonald~\cite{Wearn2016} looked into how the ethos of an educational institution correlates with the motivation of the jammers. Using separate surveys each institution, they found a difference in the motivation of the jammers at the respective sites. For example, for jammers at Staffordshire University \emph{Working with my friends} was the highest-ranking answer, while at GCU it was \emph{For the fun}

\subsection{Organisers \& Their Motivations}
\label{sec_Organisers}
Generally, we know very little about game jam organisers' motivations except for what we can infer from the overall theme of a jam. For example, there are a lot of serious game jams or game jams for a purpose, such as highlighted in the section of Purposeful Game Jams~(section~\ref{sec_PurposefulGameJams}). Unfortunately, focusing only on the reason a jam was created misses secondary and underlying motivations of organisers, such as becoming part of a community, job obligations, and so forth. There has been a bit of work related to this, such as Aurava et al.~\cite{Aurava2020} who confirmed through a study that Finnish teachers do think that game jamming can be an appropriate teaching method for soft skills such as communication and collaboration, and so student learning can be a motivation for teachers to use game jamming as an educational tool. However, the teachers also said that game jams might be best fitted for themed teaching days because of the time requirements. 

\subsection{Sharing}
\label{sec_Sharing}
The main goal of the organisers of the very first game jam~\cite{Hecker2002} was to share their creations at GDC, and even discussed open sourcing the games with the GNU General Public License (GPL)~\cite{Gay:2002:FSF}. During the first years, the Nordic Game Jam~\cite{Lai2006} took the idea of sharing further, by not just putting all the games online, but also the source code and assets. All three (game, source code, assets) were licensed as one package under the Creative Commons (CC) license. GGJ, when it modelled itself upon the Nordic Game Jam, also adopted the idea of releasing games under a CC license. The exact license and version has changed, as the CC licenses have evolved. Lai et al.~\cite{Lai2019} found a number of issues with the use of a CC license. For example, as evidenced by volunteers, a steady number of people do complain about the use of the CC license, as it is not suitable for source code~\cite{CCFAQ}. The authors instead propose a new license for game jams, \emph{The Game Jam License} which they claim rectifies the issues identified~\cite{Lai2019}.

\subsection{Theme}
\label{sec_Theme}
For many game jams, a central theme, either released ahead of time or shortly before the event begins, works to funnel the creativity and potentially level the playing field among jammers, as it makes it harder to prepare the whole or parts of a game ahead of time.

Kultima~\cite{Kultima2015a} made a grounded analysis of the work of the theme committee for GGJ 2012. She highlights how different views on the role of the theme and GGJ affected the decision process, and extracts common reasons, among committee members, to accept or reject a theme idea.

McDonald \& Moffat~\cite{McDonald2016} used sentiment analysis to determine the jammers reaction to the GGJ themes from 2010 to 2016. They did this by collecting tweets with the hashtags \#ggj10 to \#ggj16
and containing the word describing the theme of that year. The theme with the most positive sentiment was from 2012, while the year with the biggest net positive sentiment was 2016.

\section{Conclusion} 
In this paper, we have looked back on almost two decades of game jams. First, we discussed the different definitions of game jams, then going through the history of the movement, we identified different historical waves of jams, followed by a categorical taxonomy of game jams. Finally surveying literature on the subject, we identified eight areas of research.

Looking forward, with game jams already popular in the education sector and in game companies, we see this trend as continuing to grow. When examining hackathons and other participatory events with similarities to game jams, we found that there is a lack of research comparing the movements, which would be needed to identify the specificities of each type of events.

\begin{acks}
This work was funded in part by the EPSRC Centre for Doctoral Training in Intelligent Games \& Game Intelligence (IGGI) EP/L015846/1.
\end{acks}

\bibliographystyle{ACM-Reference-Format}
\bibliography{sample-base}

\end{document}